\documentclass[english]{article}
\usepackage[T1]{fontenc}
\usepackage[latin9]{inputenc}
\usepackage{geometry}
\makeatletter
\usepackage[margin=10pt,font=small,labelfont=bf,labelsep=endash]{caption}
\usepackage{microtype}
\date{}

\makeatother

\usepackage{babel}
\begin{document}

\title{Policy options for the radio detectability of Earth}

\author{Jacob Haqq-Misra\\
\\
Blue Marble Space Institute of Science\\
1001 4th Ave Suite 3201, Seattle WA 98154\\
\\
Email: jacob@bmsis.org. \\
Accepted for publication in \emph{Futures}.}
\maketitle

\section*{Abstract}

The METI risk problem refers to the uncertain outcome of sending transmissions
into space with the intention of messaging to extraterrestrial intelligence
(METI). Here, I demonstrate that this uncertainty is undecidable by
proving that that the METI risk problem reduces to the halting problem.
This implies that any proposed moratorium on METI activities cannot
be based solely on the requirement for new information. I discuss
three policy resolutions to deal with this risk ambiguity. \emph{Precautionary
malevolence} assumes that contact with ETI is likely to cause net
harm to humanity, which remains consistent with the call for a METI
moratorium, while \emph{assumed benevolence} states that METI is likely
to yield net benefits to humanity. I also propose a policy of \emph{preliminary
neutrality}, which suggests that humanity should engage in both SETI
(searching for extraterrestrial intelligence) and METI until either
one achieves its first success. \\
\\
\textbf{Keywords:} SETI; METI; radio transmission; Halting problem,
detectability

\section{Introduction: The METI risk problem}

Could sending messages into space be dangerous? A recent exchange
between Buchanan (2016) and Vakoch (2016) in the journal \emph{Nature
Physics} highlights the ongoing debate as to whether or not messaging
to extraterrestrial intelligence (METI) with electromagnetic waves
constitutes an existential risk to humanity. The potential risk associated
with METI has led some researchers to consider the possibility of
a moratorium on METI activities (Shuch \& Almár, 2007; Billingham
\& Bedford, 2011). Other analyses suggest that METI could cause either
positive or negative consequences through unknown factors that we
cannot influence (Penny 2012), while any attempts at concealing our
civilization's leakage radiation would be prohibitively burdensome
(Shostak 2013). 

I define the METI risk problem as the question: \emph{would contact
with extraterrestrial intelligence (ETI) cause net benefits or net
harm for humanity?} Because engaging in METI could possibly attract
additional attention from ETI (in addition to existing radio signals,
biosignatures, and city lights), this question is also directly relevant
for determining the value of engaging in METI. Following Haqq-Misra
et al. (2013), the value of engaging in METI, $U_{M}$, can be expressed
in terms of the probability of detection by ETI, $p_{d}$, and the
expected value of the magnitude of ETI response, $\left\langle M\right\rangle $,
as: 
\begin{equation}
U_{M}=p_{d}\left\langle M\right\rangle -C.\label{eq:METIvalue}
\end{equation}
In Eq. (\ref{eq:METIvalue}), $C$ represents the total financial
costs of engaging in METI, while the term $p_{d}\left\langle M\right\rangle $
represents the total value (positive or negative) realized from actually
conducting METI. The heart of the METI risk problem is the sign of
$\left\langle M\right\rangle $; we cannot readily predict whether
$\left\langle M\right\rangle $ should be positive or negative. This
leads to fundamental disagreements regarding whether or not to engage
in METI, as we cannot know if contact with ETI would be beneficial
($\left\langle M\right\rangle \ge0\Rightarrow U_{M}\ge0$) or harmful
($\left\langle M\right\rangle <0\Rightarrow U_{M}<0$) to humanity.

This paper demonstrates that the METI risk problem is fundamentally
undecidable, which suggests three policy options for how to proceed
(or not proceed) with METI. Each of these policy options would determine
the long-term detectability of Earth's radio spectrum to any extraterrestrial
watchers. Likewise, the same policy options should apply to any extraterrestrial
civilizations also facing the METI risk problem, which underscores
the connection between METI policy and the search for extraterrestrial
intelligence (SETI).

\section{Proof of undecidability}

Can the METI risk problem be resolved? Penny (2012) argues that METI
could be beneficial or harmful on an existential level, with no ability
to distinguish between this range of options until the actual detection
of ETI. This line of reasoning suggests that any policies for METI
today, such as a METI moratorium, cannot be based upon the requirement
for new information.

In this section, I provide a formal demonstration of the argument
advanced by Penny (2012). The purpose of this demonstration is to
highlight a key feature of the METI risk debate that is sometimes
neglected; namely, that the only information that could resolve the
METI risk problem is the actual detection of ETI. In particular, I
construct a proof by reduction to show that the METI risk problem
is undecidable. (Readers who are convinced of this statement, or find
it trivial, can skip to the next section.) A decision problem is \emph{undecidable}
if it is impossible to construct an algorithm (or Turing machine)
that can provide a yes or no answer to every question in the domain
of the problem. I accomplish this proof by demonstrating that the
METI risk problem reduces to a known undecidable problem; namely,
the halting problem.

The halting problem is a classic decision problem that can be expressed
as follows: given an arbitrary Turing machine $T_{halt}$, can we
determine (with a yes or no answer) if $T_{halt}$ will halt when
initialized with input $w$? In other words, the halting problem asks
whether or not it is possible to determine in advance if a computer
program will halt or continue running forever. The halting problem
was proved to be undecidable by Alan Turing in 1936, which implies
that any computable function that can be reduced to the halting problem
is therefore also undecidable.

To begin the proof, I first assume that there exists a decidable Turing
machine $T_{METI}$ that answers the METI risk problem when given
a set of observations $w$. (Note that $w$ represents any general
set of information, including astronomical observations, sociological
deductions, and risk analysis. $T_{METI}$ represents any possible
deductive means of using $w$ to arrive at an answer to the METI risk
problem.) I construct $T_{METI}$ as a finite-state machine with three
possible states: $q_{0}$ (initial state, $\left\langle M\right\rangle $
is unknown), $q_{1}$ ($\left\langle M\right\rangle \ge0$), and $q_{2}$
($\left\langle M\right\rangle <0$). When given input $w$, $T_{METI}$
will either remain at $q_{0}$, transition to state $q_{1}$, or transition
to state $q_{2}$. $T_{METI}$ answers \emph{yes} if state $q_{1}$is
reached and answers \emph{no} if state $q_{2}$ is reached. Likewise,
if $T_{METI}$ remains in state $q_{0}$, then this implies that input
$w$ is insufficient to reach a decision. 

I next define a Turing machine $T_{METI}^{'}$ that takes $w$ and
$T_{METI}$\textbf{ }as input. I construct $T_{METI}^{'}$ so that
$T_{METI}^{'}$ will transition to state $q$ and halt if and only
if $T_{METI}$ enters state $q_{1}$ or $q_{2}$ with input $w$.
This implies that $T_{METI}^{'}$ will halt if and only if $T_{METI}$
answers \emph{yes} or \emph{no}. I have now reduced the resolution
of the METI risk problem $T_{METI}$ to the question of whether or
not $T_{METI}^{'}$ will halt with input $w$. This is identical to
the halting problem (i.e., determining if $T_{halt}$ will halt with
input $w$). However, I assumed at the start of this proof that $T_{METI}$
is decidable, but I have reduced $T_{METI}$ to the undecidable problem
$T_{halt}$. Therefore, I have arrived at a contradiction and must
conclude that $T_{METI}$ is undecidable.

\section{Policy options}

If the METI risk problem is undecidable, then how should we as a civilization
decide whether or not to engage in METI? No information, prior to
the actual discovery of ETI, can provide an unambiguous prediction
of benefits or harm from ETI contact. In order to make any progress
on this question, we must therefore make assumptions about the terms
in Eq. (\ref{eq:METIvalue}).

\subsection{Precautionary Malevolence}

One approach, adopted by many critics of METI, is that of \emph{precautionary
malevolence}. Lacking any means of deciding the METI risk problem,
a policy of precautionary malevolence makes the assumption that $\left\langle M\right\rangle <0\Rightarrow U_{M}<0$,
which implies that we should not engage in METI. Arguments in favor
of this policy include the recognition that primitive human societies
on Earth tend to collapse, at least culturally, when directly contacted
by our modern technological civilization (Diamond, 1999). Given the
magnitude of possible risks, so the argument goes, we should refrain
from engaging in METI, or at least give careful thought to the information
content of any messages, in order to avoid increasing any risk from
ETI contact to our own civilization (Barkow 2014). 

One version of a policy of precautionary malevolence is the proposal
to enact a worldwide METI moratorium until the METI risk problem can
be resolved (Shuch \& Almár, 2007; Billingham \& Bedford, 2011). However,
I have shown above that the METI risk problem is undecidable, so no
new information (apart from the actual discovery of ETI) can resolve
this issue. Thus, a METI moratorium cannot be justified on the basis
of a learning-before-doing decision principle, because no amount of
exploration or research can resolve an undecidable problem. Korbitz
(2014) has similarly discussed the limited applicability of the precautionary
principle to defend a METI moratorium. The only possible learning
that could resolve this ambiguity is if SETI actually succeeds in
discovering a signal from ETI\textemdash in which case the decision
to lift the moratorium would become a matter of debate. Thus, a METI
moratorium on the basis of learning-before-doing would effectively
prohibit METI activities entirely until SETI first succeeds.

A modified version of precautionary malevolence is the idea that we
should wait until we are ``more evolved'' as a species before we
think about engaging in METI. This position presumes that human civilization
currently lacks the moral capacity to engage in contact with ETI,
but future humans (or transhumans) may evolve greater sensibilities
than we know now. This version of precautionary malevolence would
even suggest that we should refrain from engaging in METI even if
SETI succeeds\textemdash instead, we should wait until a more distant
time in the future when we have evolved as a species. Although I cannot
predict the future of human ethics, I will assume that any criterion
based upon human evolution must require at least hundreds to thousands
of years of civilizational development in order for culture and biology
to significantly change from today. If this is the case, and if we
assume a policy of precautionary malevolence, then it becomes fair
to ask the question: why engage in SETI at all? If we have no plans
of replying at all until a distant and unspecified point in the future,
then why even bother searching? If human ethics is the limiting factor,
then perhaps we should abandon both SETI and METI and focus instead
on improving our civilizational values.

\subsection{Assumed Benevolence}

Another approach, adopted by many advocates of METI, is that of \emph{assumed
benevolence}. Lacking any means of deciding the METI risk problem,
a policy of assumed benevolence makes the assumption that $\left\langle M\right\rangle \ge0\Rightarrow U_{M}>0$,
which implies that we should consider engaging in METI now. The primary
argument in favor of this policy is that any advanced ETI should already
have detected our technological civilization, regardless of whether
or not we engage in METI. In addition to the radio leakage that has
been emanating from Earth for nearly one hundred years (Scheffer,
2004), other tell-tale signs of intelligent life on Earth include
our city lights (Loeb \& Turner, 2012) and the spectral signature
of our atmosphere (Schneider et al., 2010). Any possible harm that
ETI could cause us, so the argument goes, should have already occurred,
so we need not worry about any catastrophic threats that could result
from METI. Furthermore, the mere discovery of ETI is presumed to be
beneficial (if only for scientific and philosophical purposes), which
leads toward the conclusion that any success that arises from METI
would provide net benefits. Some optimists further imagine that contact
with ETI could initiate an interstellar conversation that would bestow
us with advanced knowledge (e.g., Sagan 1973) or even welcome us into
a galactic club (Bracewell, 1976), bringing about tremendous benefits
that would otherwise have been impossible.

A particular version of assumed benevolence examines the possibility
of technological asymmetry between us and ETI (Vakoch, 2011). Any
ETI that we might contact will invariably be more advanced than us
(see, e.g., Baum et al., 2011), which suggests that we might make
more widespread use of radio technology than them. By this argument,
SETI may never succeed because advanced ETI neither transmit unintentional
leakage nor intentional transmissions in our direction. This technological
asymmetry suggests that the burden of transmission falls upon the
younger, less advanced, civilization. Given that ETI should already
know about our existence, our current technological position may place
us in an ideal situation to initiate contact with nearby, non-malevolent,
ETI.

A policy of assumed benevolence suggests that METI should be beneficial
in general, so we should start thinking about actively transmitting
for METI as we continue passively exploring through SETI observations.
However, assumed benevolence does not necessarily suggest that we
should transmit at all costs; METI is expensive, after all, and the
costs of engaging in METI must be weighed against the (unknown) magnitude
of benefits that would arise from contact with ETI (i.e., does $p_{d}\left\langle M\right\rangle \ge C$
in Eq. (\ref{eq:METIvalue})?). A policy of assumed benevolence is
inconsistent with the idea of a METI moratorium, although this policy
also does not necessarily imply that we must start transmitting today.
For example, the cost of METI will decrease significantly in the coming
decades when missions such as TESS, PLATO, and JWST begin to characterize
nearby exoplanets, which will provide a much better target list for
both SETI and METI.

\subsection{Preliminary Neutrality}

A third approach to the METI risk problem is a policy of \emph{preliminary
neutrality}. Lacking any means of deciding the METI risk problem,
a policy of preliminary neutrality makes the assumption that $\left\langle M\right\rangle =0\Rightarrow U_{M}=-C$.
This policy is takes no position on the likely behavior, social structure,
or technological capabilities of ETI. This implies that there is no
significant reason to engage in METI, while the only reason to oppose
METI is based upon cost limitations. Evidence for preliminary neutrality
include the lack of response by ETI to the handful of METI attempts
that could have succeeded by now. 

The policy of preliminary neutrality implies that the value of METI
is simply the cost of transmitting. This policy can be used to argue
in favor of prioritizing relevant public resources for other purposes,
such as SETI or exoplanet characterization. However, this policy is
also consistent with the desires of individuals, private corporations,
and wealthy sponsors to engage in METI at their own expense (for whom
$C$ is an acceptable sunk cost). 

Preliminary neutrality remains inconsistent with the idea of a METI
moratorium, although it cannot make a strong argument for the prioritization
of METI. Under a policy of preliminary neutrality, activities that
could potentially lead to the discovery of ETI\textemdash such as
any SETI or METI programs, and possibly exoplanet characterization\textemdash should
continue unabated as funding and enthusiasm permit \emph{until the
actual discovery of ETI}. Once this discovery occurs, all METI programs
should be suspended immediately to allow for global discussion of
this unprecedented result. Preliminary neutrality recognizes the possibility
that speculative endeavors like METI and SETI are probably unlikely
to succeed, so there is no reason to prohibit either of these activities
until direct evidence shows otherwise.

\section{Conclusion: Implications for detectability}

I have demonstrated that the METI risk problem is undecidable. That
is, the question ``would contact with ETI cause net benefits or net
harm for humanity?'' cannot be unambiguously resolved prior to the
actual detection of ETI. Based on this proof, three major policy options
remain:
\begin{enumerate}
\item Precautionary malevolence: contact with ETI is likely to cause net
harm to humanity, so we should not engage in METI until SETI succeeds.
\item Assumed benevolence: contact with ETI is likely to yield net benefits
to humanity, so we should begin engaging in cost-effective METI along
with SETI.
\item Preliminary neutrality: contact with ETI is unlikely to occur, so
we may as well continue with SETI and METI until one of them succeeds.
\end{enumerate}
Precautionary malevolence, particularly if it leads to a METI moratorium,
would decrease Earth's long-term detectability in order to mitigate
potential risk. An extreme form of such a moratorium would consider
not just radio transmissions, but also optical evidence of civilization
and even the presence of spectral technosignatures in Earth's atmosphere.
Such a policy would decrease the likelihood of Earth's detection by
ETI astronomers, and may also correlate to a decreased probability
of success for SETI. Under an active policy of precautionary malevolence,
future Earth would be intentionally much less detectable than Earth
today.

Assumed benevolence encourages engagement in METI today as a supplement
to SETI, which would increase the overall strength of radio (and perhaps
optical) transmissions from Earth. Most METI attempts today target
individual star systems, particularly those suspected to harbor habitable
planets, which can increase Earth's detectability in the direction
of transmission. Under a policy of assumed benevolence, future Earth
would be intentionally more detectable than Earth today, although
such an increase would not be uniform in all directions or across
all wavelengths.

Preliminary neutrality acknowledges that SETI and METI are both unlikely
to succeed. Human civilization faces a wide range of catastrophic
and existential threats, none of which require intervention by ETI,
so resources should be prioritized toward ensuring a long-term sustainable
future. Under a policy of preliminary neutrality, the detectability
of future Earth would remain approximately the same as today. METI,
as well as SETI, would remain viable under preliminary neutrality,
with scales limited only by the availability of resources. Preliminary
neutrality acknowledges that SETI and METI are both extremely low-probability
gambles, but the success of either one would lead to a transformational
jackpot. 

\section*{Acknowledgments}

This research did not receive any specific grant from funding agencies
in the public, commercial, or not-for-profit sectors.

\section*{References}

Barkow, J. H. (2014). Eliciting Altruism While Avoiding Xenophobia:
A Thought Experiment. In Extraterrestrial Altruism (pp. 37-48). Springer,
Berlin, Heidelberg.\\
\\
Baum, S. D., Haqq-Misra, J. D., \& Domagal-Goldman, S. D. (2011).
Would contact with extraterrestrials benefit or harm humanity? A scenario
analysis. Acta Astronautica, 68(11-12), 2114\textendash 2129.\\
\\
Billingham, J., \& Benford, J. (2011). Costs and difficulties of large-scale'messaging',
and the need for international debate on potential risks. arXiv:1102.1938.\\
\\
Bracewell, R. N. (1976). The galactic club: Intelligent life in outer
space. San Francisco, WH Freeman and Co.\\
\\
Buchanan (2016). Searching for trouble? \emph{Nature Physics} 12:
720.\\
\\
Diamond, J. (1999). To whom it may concern. New York Times Magazine,
5, 68-71.\\
\\
Loeb, A., \& Turner, E. L. (2012). Detection technique for artificially
illuminated objects in the outer solar system and beyond. Astrobiology,
12(4), 290-294.\\
\\
Haqq-Misra, J., Busch, M. W., Som, S. M., \& Baum, S. D. (2013). The
benefits and harm of transmitting into space. Space Policy, 29(1),
40\textendash 48.\\
\\
Korbitz, A. (2014). The precautionary principle: Egoism, altruism,
and the active SETI debate. In Extraterrestrial Altruism (pp. 111-127).
Springer, Berlin, Heidelberg.\\
\\
Penny, A. (2012). Transmitting (and listening) may be good (or bad).
Acta Astronautica, 78, 69-71.\\
\\
Sagan, C. (1973). Communication with extraterrestrial intelligence
(CETI). Communication With Extraterrestrial Intelligence.\\
\\
Scheffer, L. (2004). Aliens can watch \textquotedblleft I Love Lucy.\textquotedblright{}
Contact in Context, 2(1), lucy.pdf.\\
\\
Schneider, J., Léger, A., Fridlund, M., et al. (2010). The far future
of exoplanet direct characterization. Astrobiology, 10(1), 121-126.\\
\\
Shostak, S. (2013). Are transmissions to space dangerous?. International
Journal of Astrobiology, 12(1), 17-20.\\
\\
Shuch, H. P., \& Almár, I. (2007). Shouting in the jungle: the SETI
transmission debate. Journal of the British Interplanetary Society,
60(4), 142-146.\\
 \\
Vakoch, D. A. (2011). Asymmetry in Active SETI: A case for transmissions
from Earth. Acta Astronautica, 68(3-4), 476\textendash 488.\\
\\
Vakoch (2016). In defence of METI. \emph{Nature Physics}, 12: 890.
\end{document}